# Van der Waals magnetic-electrode transfer for two-dimensional spintronic devices

*Zhongzhong Luo[1,3], Zhihao Yu[3,4], Xiangqian Lu[5], Wei Niu[6], Yao Yu[4], Yu Yao[7], Fuguo Tian[7], Chee Leong Tan[4], Huabin Sun[3,4], Li Gao[6,7], Wei Qin[5]\*, Yong Xu[3,4]\*, Qiang Zhao[1,7]\* and Xiang-Xiang Song[2,8]\**

[1]College of Electronic and Optical Engineering & College of Flexible Electronics (Future Technology), State Key Laboratory of Organic Electronics and Information Displays, Nanjing University of Posts and Telecommunications, Nanjing 210023, China

[2]CAS Key Laboratory of Quantum Information, University of Science and Technology of China, Hefei, 230026, China

[3]Guangdong Greater Bay Area Institute of Integrated Circuit and System, Guangzhou 510535, China

[4]College of Integrated Circuit Science and Engineering, Nanjing University of Posts and Telecommunications, Nanjing 210023, China

[5]School of Physics, State Key Laboratory of Crystal Materials, Shandong University, Jinan 250100, China

[6]School of Science, Nanjing University of Posts and Telecommunications, Nanjing 210023, China

[7]Institute of Advanced Materials, Nanjing University of Posts and Telecommunications, Nanjing 210023, China

[8]Suzhou Institute for Advanced Research, University of Science and Technology of China Suzhou, 215123, China

\* Correspondence should be addressed to X.-X.S. (songxx90@ustc.edu.cn), Q.Z. (iamqzhao@njupt.edu.cn), Y.X. (xuyong@njupt.edu.cn) or W.Q. (wqin@sdu.edu.cn)

## ABSTRACT

Two-dimensional (2D) materials are promising candidates for spintronic applications. Maintaining their atomically-smooth interfaces during integrating ferromagnetic (FM) electrodes is crucial since conventional metal deposition tends to induce defects at the interfaces. Meanwhile, the difficulties in picking up FM metals with strong adhesion, and in achieving conductance match between FM electrodes and spin transport channels make it challenging to fabricate high-quality 2D spintronic devices using metal transfer techniques. Here, we report a solvent-free magnetic-electrode transfer technique that employs a graphene layer to assist in transferring FM metals. It also serves as part of the FM electrode after transfer for optimizing spin injection, enabling realization of spin valves with excellent performance based on various 2D materials. In addition to two-terminal devices, we demonstrate that the technique is applicable for four-terminal spin valves with non-local geometry. Our results provide a promising future of realizing 2D spintronic applications using the developed magnetic-electrode transfer technique.

## TOC

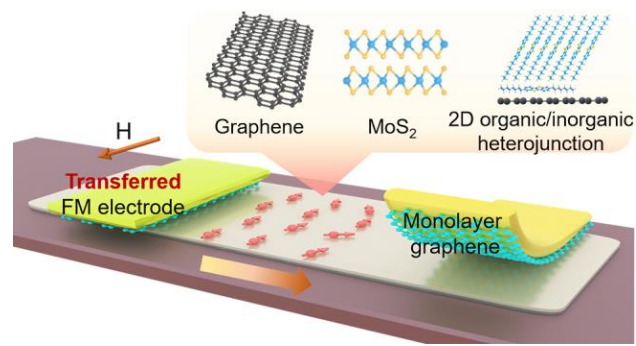

## KEYWORDS

magnetic-electrode transfer technique, spin valves, solvent-free, two-dimensional materials, high-quality interfaces, spintronics

Since the discovery of graphene, a variety of two-dimensional (2D) materials have attracted significant research interests for spintronic applications.[1-8] For example, graphene is recognized as a promising material for spin channel due to its long spin diffusion length $l_{sf}$ and high carrier mobility up to room temperature.[9-11] The intrinsic spin-orbit coupling and the bandgap in semiconducting transition metal dichalcogenides provide powerful knobs for electrically controlling the device magnetoresistance (MR) and manipulating spins they host.[12-15] Besides, owing to the weak spin-orbit coupling and hyperfine interaction,[16-18] 2D organic materials are expected to carry spins with long life time, which are essential for spintronic applications.

To unlock the great potentials that 2D materials host for spintronic devices, it is highly desirable to maintain high-quality interfaces during integrating ferromagnetic (FM) metals for spin injection.[19-21] This is technically challenging because of two reasons. One is conventional fabrication processes, such as electron beam lithography and metal deposition, tend to create defects, strain, disorder and metal diffusion at metal/2D material interfaces.[22] The other is the difficulties in achieving good conductance match between FM electrodes and spin transport channels.[23] Recently, van der Waals (vdW) metal transfer techniques have been developed and are demonstrated to be powerful for realizing atomically-smooth non-magnetic metal/2D semiconductor interfaces.[22, 24-27] However, their capability in spintronic applications[28, 29] remains highly unexplored, since 1) FM metals are usually difficult to be picked up due to their strong adhesion to substrates; 2) the obtained tunnel barriers at the contacts are usually too transparent for spin injection which need to be adjusted.

Here, we report a graphene-assisted vdW magnetic-electrode transfer technique that enables high-quality interfaces between different FM electrodes and various 2D materials. A monolayer graphene (MLG) is employed, not only to assist in peeling off the prepatterned FM electrodes from a sacrificial substrate, but also to serve as an inserting layer to modulate the contact resistance for optimizing spin injection. Based on the developed solvent-free magnetic-electrode transfer technique, atomically-smooth interfaces are achieved, enabling high-quality lateral spin valve devices equipped with gate electrodes for electrical tuning. We realize spin valves based on few-layer graphene (FLG), whose MR is among the best values reported in the literature. Further, we apply the technique to semiconducting $MoS_2$ to demonstrate transport of in-plane spins for ~1.4 μm and a maximum |MR| as large as ~20%, both of which are larger than previous results. According to our analysis, the tunneling interface resistance $R_T$ is comparable with the spin resistance of the channel $R_{ch}$ in these devices, which is beneficial for spin injection. Finally, we show that the technique is applicable for realizing gate tunable spin valves based on fragile 2D organic/inorganic heterojunctions, and four-terminal spin valves for non-local spin detections. Our work unveils the powerful capability of FM electrode integrations using the developed vdW metal transfer technique, offering a promising future of 2D material-based spintronic applications.

**Van der Waals magnetic-electrode transfer technique.** Figure 1a shows the schematic of the device investigated in this work. It consists of a 2D spin transport channel (made of graphene, $MoS_2$, or 2D organic/inorganic heterojunctions) and two transferred FM electrodes ($Fe_{20}Ni_{80}$ and Co, respectively), serving as a two-terminal lateral spin valve. The FM electrodes are made of three layers of MLG, FM metal (Co or $Fe_{20}Ni_{80}$), and Au, which are deposited in sequence on a sacrificial substrate of $SiO_2$/Si. Taking advantage of the weak adhesion between MLG and $SiO_2$, the FM electrodes are easily peeled off from the sacrificial substrate, and then are placed onto different target 2D materials (see Figure 1b). It is worth noting that the MLG layer is detached from the substrate (see Figure 1c), which is different from the previously reported graphene-assisted metal transfer printing technology.[26] It protects FM metals from oxidation during the transfer process, and serves as part of the electrode to modulate the contact resistance for optimizing spin injection from FM metals to 2D materials. The vdW interaction between MLG and the target 2D material tends to result in a closely-attached interface, which is beneficial for spin injection and detection. Using this solvent-free technique, we can non-destructively place FM electrodes on various 2D materials to fabricate high-quality spin valve devices. More information can be found in Note S1 in Supporting Information.

The pristine quality of FM electrode/2D material interfaces is revealed by high-resolution transmission electron microscopy (TEM). Figure 1d (Figure 1f) shows a cross-sectional TEM image cutting underneath the $Fe_{20}Ni_{80}$ electrode, which is transferred onto FLG ($MoS_2$). The corresponding energy-dispersive x-ray spectroscopy (EDS) images are shown in Figure 2e (Figure 2g), from which an atomically-smooth interface between FM metal and FLG ($MoS_2$) is clearly resolved. No obvious diffusion of Ni and Fe atoms is found in the FLG ($MoS_2$) layer. Based on such damage-free interfaces, high-quality spin valve devices with effective spin injection are expected.

**Graphene-based spin valves.** Graphene has attracted significant research interests since its long $l_{sf}$ and high carrier mobility are essential for spin transport.[10, 14, 30] Figure 2a shows the cross-sectional schematic of the FLG-based device. A drain-source bias ($V_{ds}$) is applied between the two FM electrodes to generate a current of $I_d$ to be measured. The heavily-doped Si substrate acts as a back gate where a voltage of $V_{gs}$ can be applied (see Figure S2).

Figure 2b shows typical resistance curves when sweeping in-plane magnetic field at 1.5 K. Magnetizations of two FM electrodes (indicated by the arrows) flip as the field changes, resulting in different resistances. Here the MR ratio is defined as MR = ($R_H$ - $R_P$)/$R_P$, where $R_H$ is the magnetic field dependent resistance and $R_P$ is the resistance corresponding to the parallel configuration. The measured |MR| can be modulated by $V_{ds}$ and $V_{gs}$, respectively (see Figure S3). As shown in Figure 2d, |MR| reaches a maximum value of 8.6% at $V_{ds}$ = -0.1 mV and decreases at higher |$V_{ds}$|, which can be understood as excitation of magnons, band bending, and involvement of interface states at high voltages.[31] Meanwhile, as shown in Figure 2e, |MR| exhibits a maximum value near the charge neutrality point at $V_{gs}$ = 0 V (see the transfer characteristic curve shown in Figure S2b). Away from the charge neutrality point, the channel conductance $\sigma_{ch}$ increases, while |MR| decreases. This inverse scaling of MR with $\sigma_{ch}$ is associated with spin injection process.[32] According to the one-dimensional drift/diffusion theory of spin transport,[33] MR scales with $1/\sigma_{ch}$ for tunneling contacts, while scales with $\sigma_{ch}$ for transparent contacts. Therefore, our results indicate that tunneling behavior is dominant at the contacts, which is consistent with the measured nonlinear $I_d$-$V_{ds}$ curve (see Figure S2a) and the increased resistance at lower temperature (see Figure S4). This tunneling contact behavior is favorable for spin injection.

We use a standard drift/diffusion model[34, 35] to estimate the spin polarization $r_G$ at the FM electrode/FLG interface (see Note S2 in Supporting Information). Previous studies reveal that $l_{sf}$ of FLG on SiO$_2$ substrate without any treatments is about several micrometers.[30] In this range, the spin polarization is estimated to be larger than 30% when $V_{ds}$ = -0.1 mV (see Figure 2f). For comparison, a theoretical value of 33% is predicted in ideally lattice-matched FM metal/MLG/FM metal structures, resulting from the K-point spin filtering mechanism.[36] This suggests that the non-destructive vdW magnetic-electrode transfer technique provides an ideal spin injection interface.

More importantly, we investigate the spin valve effect at elevated temperature and observe that |MR| persists up to 300 K (see Figure 2c). As shown in Figure 2g, we benchmark our data with previous results,[1, 35, 37-41] obtained from similar two-terminal lateral spin valves based on FLG. A

trend of |MR| decrease with increased temperature is observed, which can be explained by the spin wave excitation model.[42] It is demonstrated that |MR| of our devices are among the best values at most of temperature ranges. This indicates the powerful capability of the developed technique in achieving effective spin polarization.

**MoS$_2$-based spin valves.** The intrinsic spin-orbit coupling in monolayer semiconducting MoS$_2$ offers a powerful knob for electrical control of spins.[14, 43] However, it also leads to quick relaxation of in-plane spins due to the D'yakonov-Perel' (DP) mechanism.[44, 45] Previous studies have suggested that transition metal dichalcogenides with even layers recover the inversion symmetry, effectively suppressing the DP spin relaxation.[13] As a consequence, lateral spin valves based on even-layer MoS$_2$ are expected to carry spins with slower spin relaxation and longer $l_{sf}$.

Here, we investigate devices based on bilayer MoS$_2$ (see Figure 3a) and focus on their performance at low temperature. From the dc transfer curve (see Figure S5a), the two-terminal field-effect mobility is estimated to be ~30 cm$^2$V$^{-1}$s$^{-1}$ at 0.3 K. The relatively low value is probably due to tunneling-dominant behaviors at the contacts, which is revealed by the quasi-symmetric nonlinearity in output curves (see Figure S5b). Then we use the lock-in technique to study spin transport behaviors. As shown in Figures 3b and 3c, pronounced spin valve effect is observed, which can be strongly modulated by gate voltage $V_{gs}$. Figure 3d shows maximum measurable |MR| as a function of $V_{gs}$. It increases first and then decreases, reaching a maximum value of 22.1% at $V_{gs} = 34$ V.

This behavior is similar to the previous observation, which is understood as manipulating the ratio between $R_T$ and $R_{ch}$ (Ref. 13). Figure 3e shows calculated MR as a function of $R_T/R_{ch}$ based on Equation S1 and experimental parameters. MR reaches its maximum when $R_T \approx R_{ch}$, which corresponds to a perfect balance between the spin injection rate and the spin relaxation rate. A larger $R_T$ decreases the spin injection rate, while a larger $R_{ch}$ induces a spin-backflow process causing the relaxation of spin within the FM electrodes. In MoS$_2$ devices, $R_T$ is affected by the Schottky barriers at the contacts, and $R_{ch}$ depends on the resistivity $\rho$ thus the carrier density as well as the mobility. All of these parameters are modulated by $V_{gs}$ (see Figure S5). Therefore, MR exhibits a peak upon varying $V_{gs}$. Based on this physical picture, $R_T$ is estimated to be comparable with $R_{ch}$ at $V_{gs} = 34$ V. Assuming $R_T=R_{ch}$, we obtain $\rho \sim 10^5$ Ω in our bilayer MoS$_2$ devices, which is consistent with reported values at low temperature.[13]

It is worth noting that the peak |MR| of our device is as large as ~20%, which is much larger than ~1% reported in Ref. 13. Using a similar way, we roughly estimate that $l_{sf}$ in our device is as large as 1.25 μm, which is about 5 times larger than the previous result.[13] This increase may result from the solvent-free feature of the technique developed in this work, since exposing 2D materials to solvents is demonstrated to cause additional spin scattering.[46] Indeed, we observe clear spin valve effect in an additional MoS$_2$ device with a channel length of ~1.4 μm (see Figure S6). Our results suggest in-plane spins can transport over a microscale distance in bilayer MoS$_2$, offering a promising future of gate-tunable spintronic devices based on semiconducting MoS$_2$.

**Theoretical analysis of FLG- and MoS$_2$-based devices.** According to our drift/diffusion model, $r_G$ is estimated to be ~30%. However, such a value cannot account for the large |MR| measured in MoS$_2$-based devices, no matter what $l_{sf}$ of MoS$_2$ is chosen. This strongly hints at a larger spin polarization $r_M$ at the FM electrode/MoS$_2$ interface (similar to Ref. 13, we use $r_M$=50% for obtaining Figure 3e, which corresponds to the maximum spin polarization of Fe$_{20}$Ni$_{80}$). To better understand this, we perform density functional theory (DFT) calculations for Fe-Ni/MLG/FLG and Fe-Ni/MLG/2L-MoS$_2$ stacks (see Figures 4a and 4b, respectively). Note that maximum |MR| of FLG-based devices (8.6%) appears at $V_{gs}$= 0 V, while |MR| of MoS$_2$-based devices peaks (22.1%) at $V_{gs}$= 34 V (corresponding to an electron density of 3×10$^{12}$ cm$^{-2}$). Therefore, we calculate spin-dependent densities of state (DOSs) of FLG- and MoS$_2$-based stacks at both pristine case and at the doping level of $n_s$=3×10$^{12}$ cm$^{-2}$ (see Figures 4c and 4d, respectively). Spin polarizations at the interfaces are estimated[19] as $P_{G/M} = \left|\frac{n_\uparrow - n_\downarrow}{n_\uparrow + n_\downarrow}\right|$, where G (M) represents FLG (MoS$_2$), and $n_\uparrow$ ($n_\downarrow$) is the spin-up (spin-down) DOS at the Fermi level (dashed lines in Figures 4c and 4d). Although for the pristine case, $P_G$ is almost equal to $P_M$, their difference increases dramatically upon electrical doping. This is because that the semiconducting nature of MoS$_2$ enables effective tuning of the Fermi level, while the gating effect of FLG is rather limited. As a result, $P_M$ ($n_s$=3×10$^{12}$ cm$^{-2}$) is larger than $P_G$ (pristine) under the experimental conditions (see Figure 4e), which agrees well with a larger $r_M$ estimated from our drift/diffusion model qualitatively. We believe this is the main reason for the larger |MR| found in our MoS$_2$-based devices.

In addition to modulating the interface spin polarization, the gating effect also strongly manipulates $R_T/R_{ch}$ in semiconducting MoS$_2$ (as discussed in Figure 3e), while has limited influence in FLG. We can calculate $R_T$ and $R_{ch}$ in our FLG-based devices to examine the ratio of $R_T/R_{ch}$ based on Equation S1 and parameters of $\rho \sim 589$ Ω (measured from an additional four-terminal device as shown in our previous work[28]), $r_G \sim 30\%$, and $l_{sf} \sim 9$ μm (determined by the measured |MR|=8.6% from Figure 2f). Interestingly, $R_T$ is found to be comparable with $R_{ch}$ in our FLG-based devices (see the blue hexagon in Figure 4f), even if variations in $r_G$ are considered (see Figure S7). This good conductance match also contributes to large |MR| in our FLG-based devices, as benchmarked in Figure 2g.

Finally, we would like to briefly comment on the role of MLG as an inserting layer for optimizing the spin injection, although it has been debated for more than a decade and is still not fully understood.[19, 36, 47-53] Besides the intuitive idea of using the large out-of-plane resistance of graphene to serve as a tunneling layer, theoretical proposals illustrate that matching spin-polarized bands in FM metals and the electronic states in MLG results in a phenomenon known as "K-point spin filtering".[36, 53] Recently, it is demonstrated that the proximity effect cannot be neglected, which results in strong interfacial hybridization between MLG and the FM metal[19] accounting for a larger measured MR. In addition, metal-induced doping and exchange-proximity induced spin polarization in MLG also contribute to the spin injection.[50] In our devices, although experimental evidence hints at pronounced tunneling and good conductance match at the contacts, DFT calculations suggest interfacial hybridization is also relevant (see Figures 4a and 4b). Further experiments can be expected on nanojunctions realized by transferring our multilayer magnetic-electrodes onto monocrystalline FM metals with controlled twist angles[19] to investigate the two aforementioned mechanisms in more details. Nevertheless, our results demonstrate the developed graphene-assisted vdW magnetic-electrode transfer technique is powerful for obtaining high-quality FM electrode/2D material interfaces, which are responsible for effective spin injection.

**Spin valves based on 2D organic/inorganic heterojunctions.** Spins in 2D organic materials are expected to have long life time due to the relatively weak spin-orbit coupling and hyperfine interaction.[54, 55] However, they are fragile and are easily damaged during conventional fabrication processes. Using the developed technique, we are able to realize gate-tunable spin valve devices based on the heterojunction of bilayer $C_8$-BTBT and FLG (see Figure 5a). The FLG flake not only acts as the substrate for synthesizing the organic $C_8$-BTBT crystal, but also serves as a spin transport channel. Beyond our previous work[28], we demonstrate MR can be tuned upon varying $V_{gs}$ (see Figure 5b). We would like to point out that a symmetric sign inversion of MR is observed when the polarity of $V_{ds}$ is changed in our devices[28] (see Figure S8), which is different from previous reports.[56-59] The physical mechanism is not understood yet which invites further investigations.

**Four-terminal spin valve devices.** Besides the investigated two-terminal spin valve devices, the developed vdW magnetic-electrode transfer technique is also applicable for four-terminal devices. By sequentially transferring electrodes onto a chosen FLG flake, four-terminal devices can be fabricated (see Figure 5c), allowing non-local investigations of spin transport. Here, we use the standard lock-in technique to measure the non-local resistance $\Delta R_{NL}$ (see Figure S9). Figure 5d shows $\Delta R_{NL}$ as a function of bias current $I_{dc}$, showing a decrease as $I_{dc}$ changes from -100 µA to 100 µA. This can be attributed to the electric field-induced spin drift in the spin injection channel.[60, 61] In addition, we observe $\Delta R_{NL}$ ~ 0.04 Ω in an additional device (see Figure S10).

In conclusion, we report a graphene-assisted magnetic-electrode transfer technique that enables spin valve devices with atomically-smooth and damage-free interfaces between FM metals and various 2D materials. These include FLG devices that exhibit |MR| which is among the best values in the literature, $MoS_2$ devices with a record high |MR| of ~20% and an in-plane spin transport distance of ~1.4 µm, and hybrid devices based on 2D organic/inorganic heterojunctions that enable gate control on MR. We also demonstrate that the technique is applicable for four-terminal spin valves allowing non-local measurements. The developed solvent-free technique prevents 2D spin transport channel from contacting with solvents, which suppresses spin dephasing resulting from contaminations by solvents.[46] Making use of the diversity of 2D organic[59, 62-64], inorganic materials[2, 6, 14, 65] and their heterojunctions,[66, 67] our work provides a promising future of employing the vdW metal transfer technique for advanced 2D spintronic applications.

ASSOCIATED CONTENT

**Supporting Information**. The Supporting Information is available free of charge at XXXXX. Experimental methods, estimation of spin polarization in FLG-based devices, transfer and output characteristic curves of the devices in the main text, additional MR curves including those from additional devices, details of data points used for benchmarking, and calculation of $R_T/R_{ch}$ for FLG-based devices.

AUTHOR INFORMATION

**Corresponding Author**

*Email: (W.Q.) wqin@sdu.edu.cn.

*Email: (Y.X.) xuyong@njupt.edu.cn.

*Email: (Q.Z.) iamqzhao@njupt.edu.cn.

*Email: (X.-X.S.) songxx90@ustc.edu.cn.

**Author Contributions**

Z.L. and X.-X.S. conceived the project. Z.L. and X.-X.S. designed the experiment. Z.L., Z.Y., Y.Yu, Y.Yao, C.L.T., H.S., L.G., Y.X., Q.Z., and X.-X.S. contributed to sample preparation, characterization, device fabrication, measurements, and data analysis. Z.L. W.N. and F.T. prepared the TEM specimens. X.L and W.Q. carried out the DFT calculations. Z.L. and X.-X.S. cowrote the manuscript with inputs from other authors. All authors contributed to discussions.

**Notes**

The Authors declare no conflict of interest.

ACKNOWLEDGMENT

We would like to thank Xinran Wang (Nanjing University), Guoping Guo (University of Science and Technology of China) and Wei Han (Peking University) for the support in this work. The authors acknowledge support from the Natural Science Foundation of Jiangsu Province (Grant No. BK20220397), the National Natural Science Foundation of China (Grant Nos. 62204130, 12274397), the National Funds for Distinguished Young Scientists (Grant No. 61825503), the Natural Science Foundation of the Higher Education Institutions of Jiangsu Province (Grant No.

22KJB510010), the Science Foundation of Nanjing University of Posts and Telecommunications (Grant No. NY221116), Guangdong Province Research and Development in Key Fields from Guangdong Greater Bay Area Institute of Integrated Circuit and System (No. 2021B0101280002) and Guangzhou City Research and Development Program in Key Field (No. 20210302001)

**Figures:**

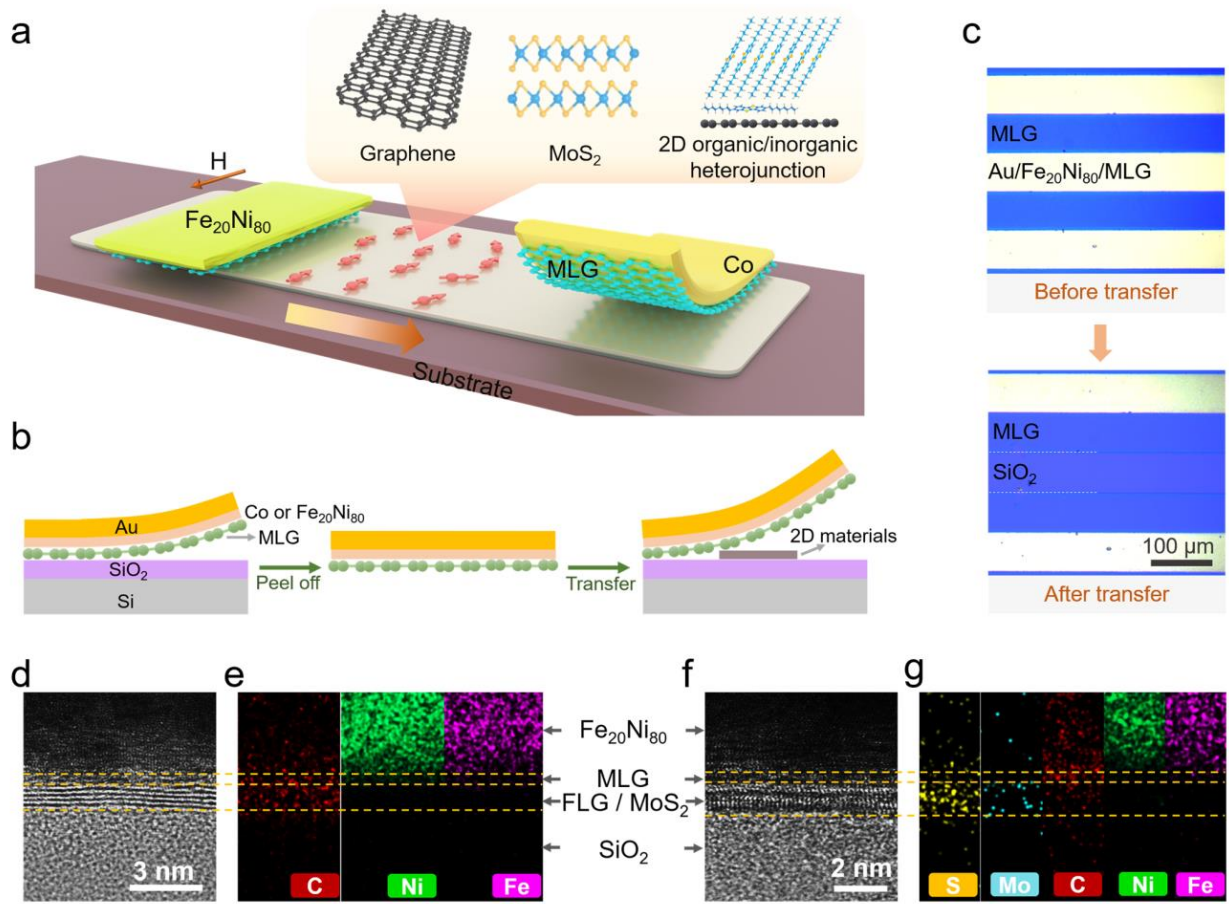

**Figure 1.** a) Schematic of a lateral spin valve device consisting of a 2D spin transport channel (made of graphene, $MoS_2$, or 2D organic/inorganic heterojunctions) and two transferred FM electrodes ($Fe_{20}Ni_{80}$ and Co, respectively). b) Schematic of the graphene-assisted vdW magnetic-electrode transfer technique for FM electrode integrations. c) Optical images of multilayer FM electrodes on the sacrificial substrate before (top) and after (bottom) electrode transfer. Edges (white dotted lines) between MLG and exposed $SiO_2$ surface indicate that MLG is detached from the sacrificial substrate and serves as part of the multilayer FM electrode. d) Cross-sectional TEM image cutting underneath the $Fe_{20}Ni_{80}$ electrode which is transferred onto FLG and e) the corresponding elemental mappings. f) Cross-sectional TEM image cutting underneath the $Fe_{20}Ni_{80}$ electrode which is transferred onto $MoS_2$ and g) the corresponding elemental mappings.

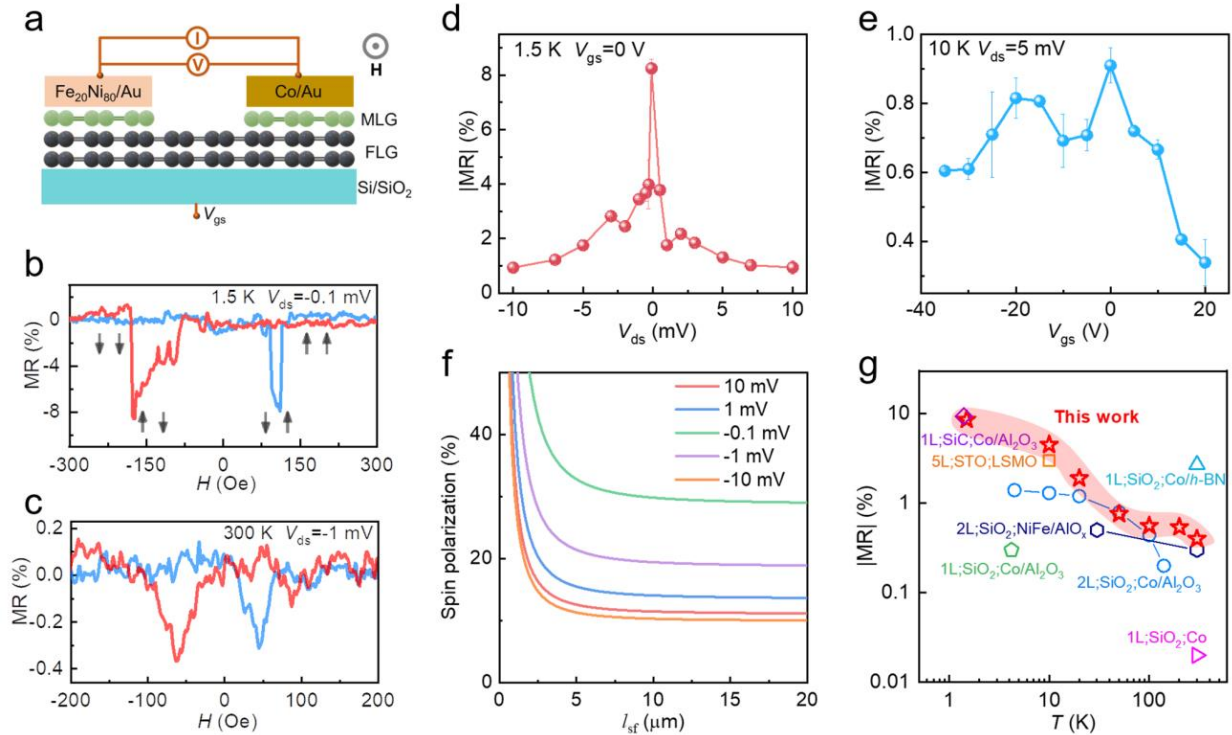

**Figure 2.** a) Schematic of a two-terminal lateral FLG-based spin valve with transferred FM electrodes. MR curves measured b) at 1.5 K, $V_{ds}$=-0.1 mV, and c) at 300 K, $V_{ds}$=-1 mV, respectively. Arrows indicate magnetizations of two FM electrodes. Extracted maximum measurable |MR| as functions of d) bias voltage, and e) gate voltage, respectively. f) Calculated spin polarization at different $l_{sf}$ based on Equation S1. Each curve corresponds to the data obtained at a fixed $V_{ds}$ shown in d). g) Benchmarking of |MR| as a function of temperature with results obtained from similar two-terminal lateral spin valves based on FLG reported in the literature. Labels mark the thickness of graphene, the substrate, and the FM electrode. Details of data points are presented in Table S1.

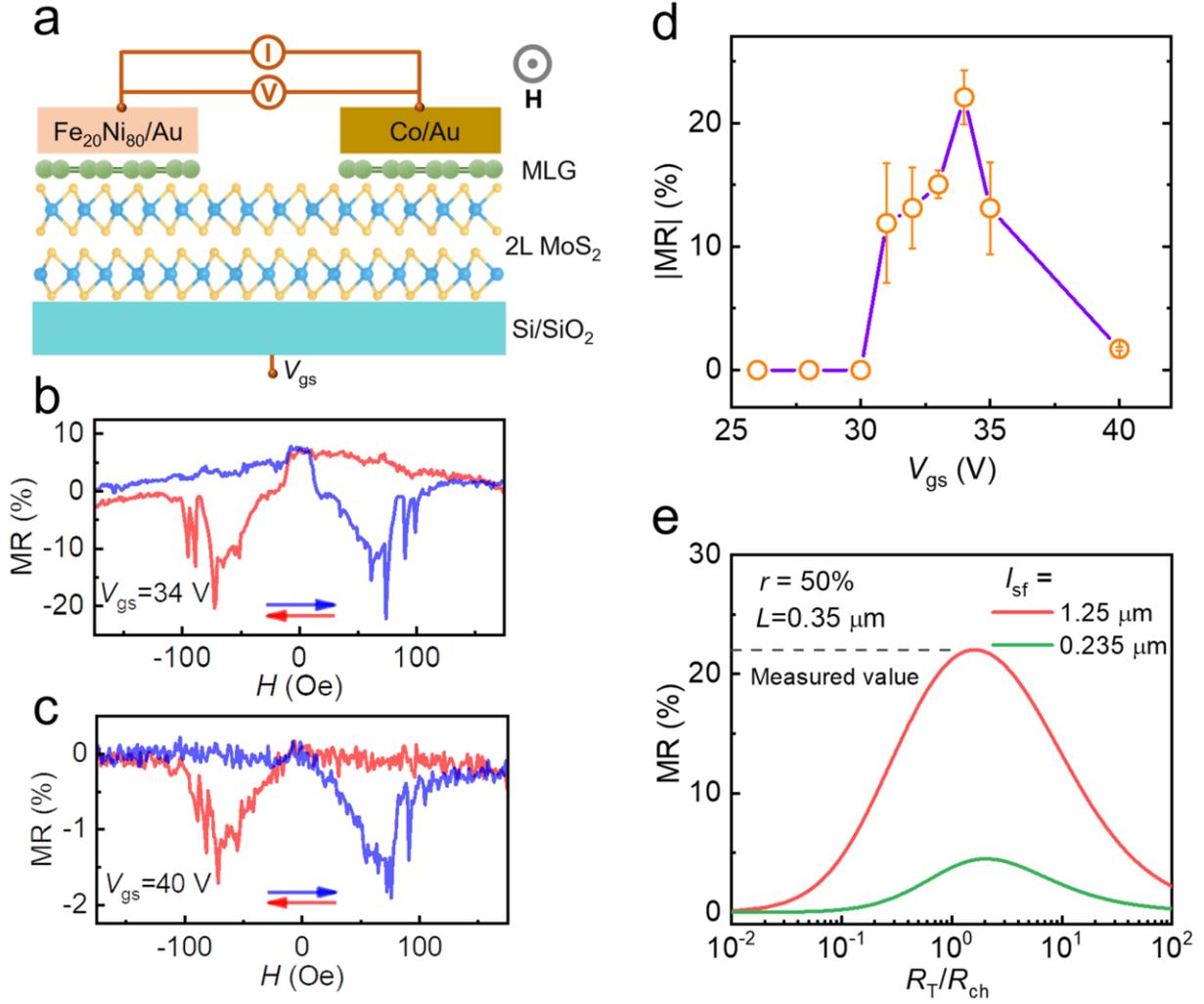

**Figure 3.** a) Schematic of a two-terminal lateral MoS$_2$-based spin valve with transferred FM electrodes. MR curves measured at 0.3 K, $V_{ds}$ = -2 mV, with b) $V_{gs}$ = 34 V, and c) $V_{gs}$ = 40 V, respectively. Arrows indicate sweeping directions of the magnetic field. The channel length and width of the tested device are 0.35 μm and 4.6 μm, respectively. d) Extracted maximum measurable |MR| as a function of gate voltage. e) Calculated MR as a function of $R_T/R_{ch}$ based on Equation S1. Assuming the interface spin polarization of ~50% (Ref. 13), corresponding to the maximum spin polarization of Fe$_{20}$Ni$_{80}$, $l_{sf}$=1.25 μm is needed to reproduce the measured maximum |MR| of 22.1% (red curve). For comparison, the green curve corresponds to the calculated result based on $l_{sf}$=0.235 μm reported previously (Ref. 13).

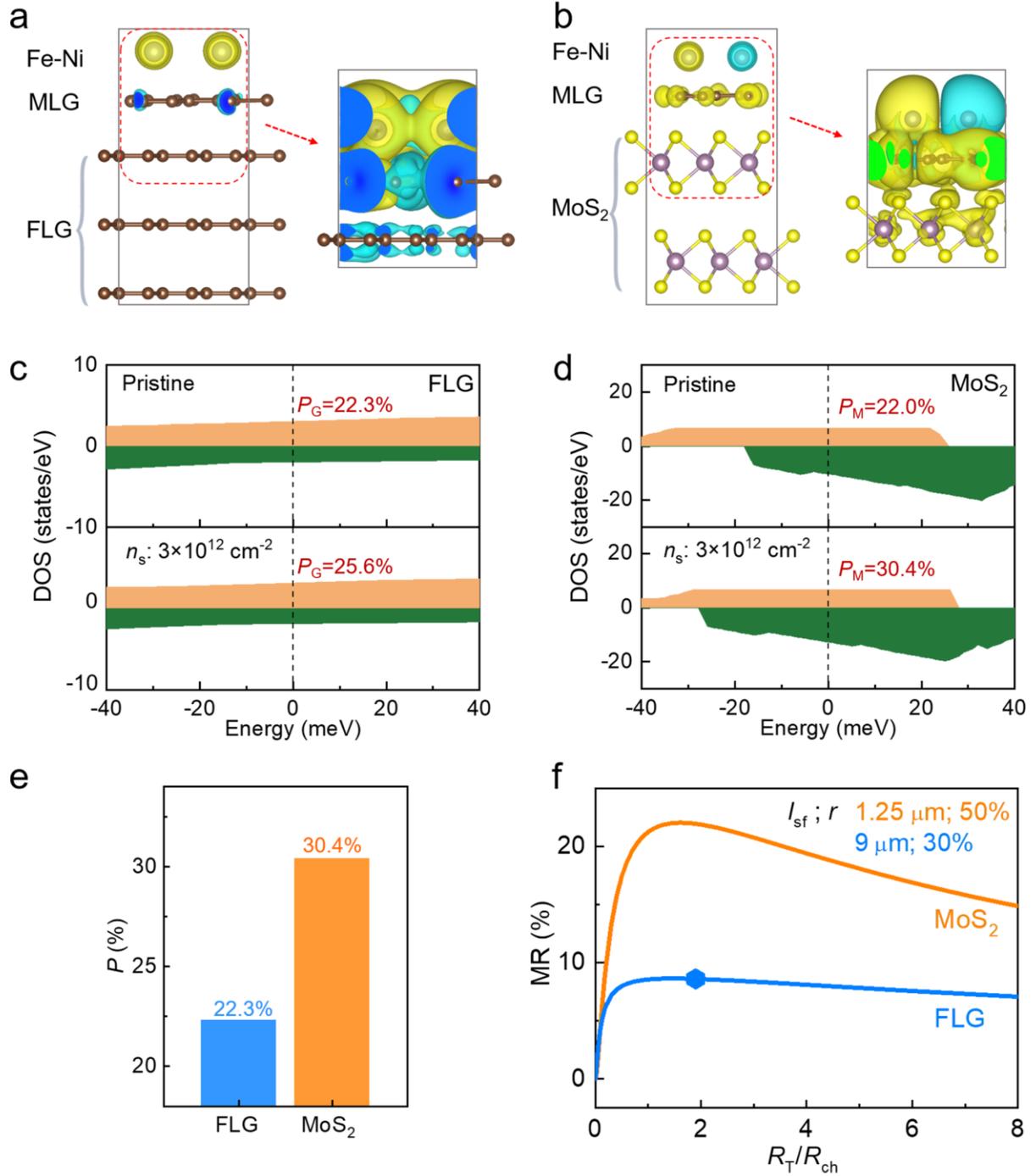

**Figure 4.** Schematic illustration of a) Fe-Ni/MLG/FLG and b) Fe-Ni/MLG/2L-MoS$_2$ stacks for DFT calculations, respectively. The black boxes mark the unit cell used in DFT calculations, which are overlaid by the spin dependent charge density distributions (cyan and yellow represent spin up and spin down, respectively). The right panels in a) and b) are zoom-in images of the red dashed boxes, highlighting spin dependent charge density distributions near interfaces. Calculated spin

dependent DOSs for c) Fe-Ni/MLG/FLG and d) Fe-Ni/MLG/2L-MoS$_2$ stacks, respectively. The upper panels in c) and d) correspond to the scenario without additional electrons induced (referred to as pristine), while the bottom panels pertain to the condition where an additional electron density of $n_s$=3×10$^{12}$ cm$^{-2}$ is introduced. $P_G$ ($P_M$) represents calculated interface spin polarization for the FLG-based (MoS$_2$-based) stack. e) Calculated $P_G$ (pristine) and $P_M$ ($n_s$=3×10$^{12}$ cm$^{-2}$) under the experimental conditions, showing a clear difference. f) Calculated MR as a function of $R_T/R_{ch}$ for FLG-based (blue) and MoS$_2$-based devices (orange) using Equation S1, respectively. The blue hexagon marks the working point of $R_T/R_{ch}$, which is calculated based on experimental parameters.

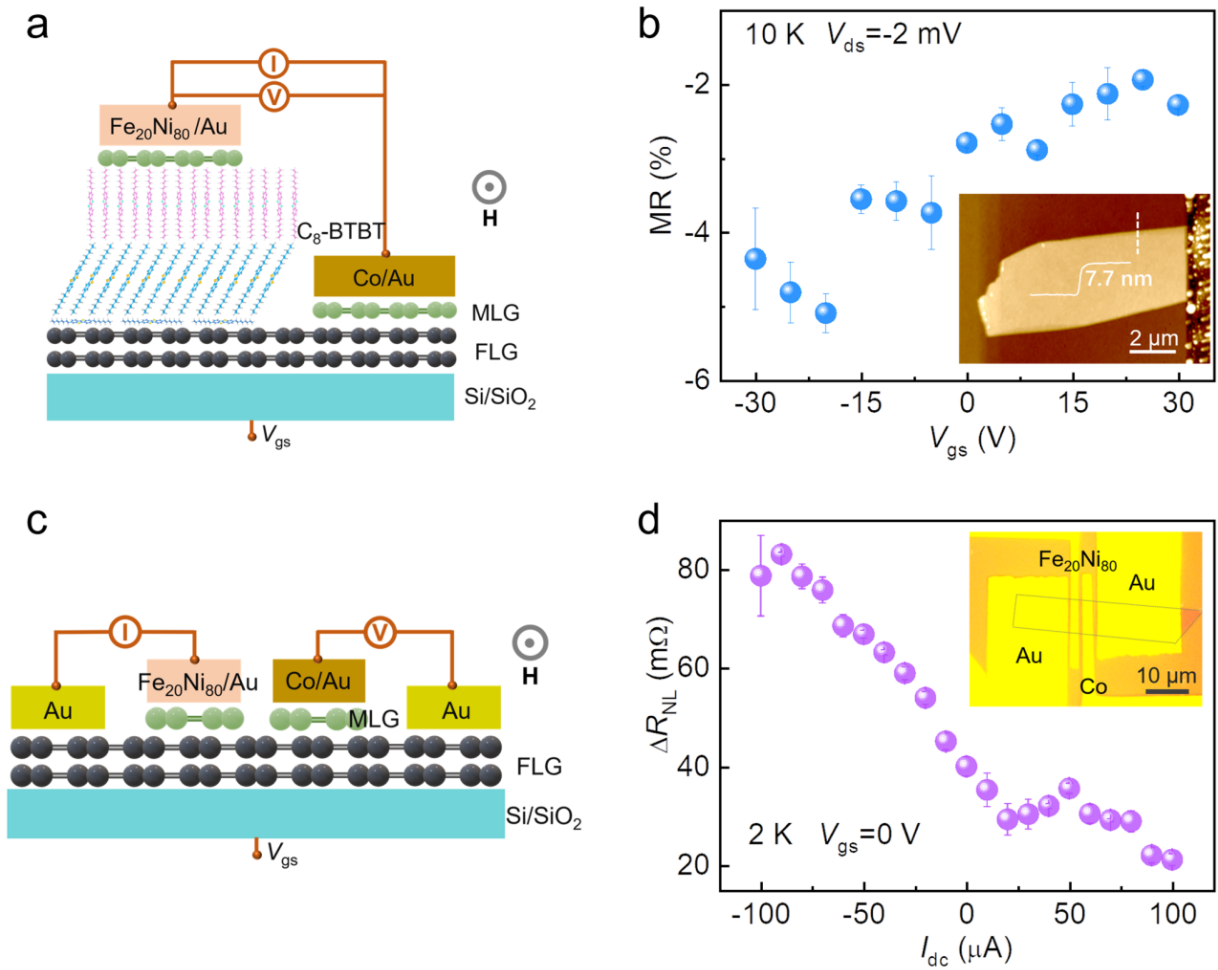

**Figure 5.** a) Schematic of a two-terminal hybrid spin valve based on 2D organic/inorganic heterojunction, where transferred FM electrodes are used for spin injection and detection. b) Extracted maximum measurable MR as a function of $V_{gs}$ obtained at $V_{ds}$ of -2 mV. The data are collected at 10 K. Inset shows an AFM image of the heterojunction of bilayer $C_8$-BTBT and FLG with the total thickness of ~7.7 nm. c) Schematic of a four-terminal FLG-based spin valve with corresponding optical image shown in the inset of d). Gray dotted lines highlight edges of the FLG channel. The device has two inner FM electrodes ($Fe_{20}Ni_{80}$ (2 μm in width) and Co (2.5 μm in width), respectively) and two outer non-magnetic electrodes. d) Non-local resistance $\Delta R_{NL}$ as a function of $I_{dc}$ measured at 2 K.